\documentclass[reprint, superscriptaddress,
 amsmath,amssymb,
 aps,
prl,showpacs,showkeys
,nofootinbib
]{revtex4-1}
\usepackage{graphicx}
\usepackage{ulem}
\usepackage{lipsum}
\usepackage{amsmath}
\usepackage{mathtools}
\usepackage{rotating}
\usepackage{tikz}
\usepackage{dcolumn}
\usepackage{xcolor}
\usepackage{amssymb}

\begin{document}

\title{Light transport through amorphous photonic materials with localization and bandgap regimes}

\author{Frank \surname{Scheffold}}
\email{frank.scheffold@unifr.ch}
\affiliation{%
Department of Physics, University of Fribourg, CH-1700 Fribourg, Switzerland
}%
\author{Jakub \surname{Haberko}}
\affiliation{%
Faculty of Physics and Applied Computer Science, AGH University of Science and Technology, Al. Mickiewicza 30, Krakow 30-059, Poland.
}%

\author{Sofia \surname{Magkiriadou}}
\affiliation{%
Department of Physics, University of Fribourg, CH-1700 Fribourg, Switzerland
}%
\author{Luis S. \surname{Froufe-P\'erez}}
\affiliation{%
Department of Physics, University of Fribourg, CH-1700 Fribourg, Switzerland
}%
\begin{abstract}
We propose a framework that unifies the description of light transmission through three-dimensional amorphous dielectric materials that exhibit both localization and a photonic bandgap. We argue that direct, coherent reflection near and in the bandgap attenuates the generation of diffuse or localized photons. Using the self-consistent theory of localization and considering the density of states of photons, we can quantitatively describe the total transmission of light for all transport regimes: transparency, light diffusion, localization, and bandgap. Comparison with numerical simulations of light transport through hyperuniform networks supports our theoretical approach.
\end{abstract}

\maketitle
Photonic bandgaps (PBG) and light localization fundamentally alter a dielectric material's wave transport properties~\cite{Joannopoulos_book,lagendijk2009fifty}. In 1987 Yablonovitch proposed that crystal lattices composed of high and low index dielectric materials can lead to forbidden propagation in certain electromagnetic frequency bands ~\cite{yablonovitch1987inhibited}. More recently, researchers demonstrated the existence of bandgaps also in two- and three-dimensional disordered, amorphous dielectrics based on numerical simulations and experiments~\cite{edagawa2008photonic, Florescu_PNAS_2009, Cao_PRA_2011, Muller_Scheffold_Adv_Opt_Mat_2014,man2013isotropic,sellers2017local,Froufe_PRL_2016}. In particular, disordered 'hyperuniform' photonic materials raised a lot of attention~\cite{torquato2015ensemble}. Several groups showed that these materials could exhibit isotropic complete photonic bandgaps nearly as wide as the corresponding crystal structure~\cite{edagawa2008photonic,Florescu_PNAS_2009,Cao_PRA_2011}. Bandgaps in amorphous dielectrics have renewed interest in strong Anderson localization (SAL) of light and other transport regimes in those materials. We, with others, proposed a transport phase diagram to organize numerical and experimental data~\cite{froufe2017transport,aubry2020experimental,sgrignuoli2021localization}.  
However, recent three-dimensional finite-difference time-domain (FDTD) simulations have also shown that existing theoretical models cannot describe the transition between the localization and bandgap regimes~\cite{haberko2020transition}, calling for a new and improved theoretical approach. 
In this letter, we introduce a theoretical model based on the self-consistent theory of localization (SC-theory) of a semi-infinite medium~\cite{wolfle2010self,cherroret2010transverse,van2000reflection}, together with an exponential direct reflection coefficient. We show that our model is capable of describing the transmission of light through amorphous photonic materials over the entire range of frequencies, encompassing all transport regimes.
\newline {\it{Ballistic and diffuse transport of light.--}} The transmission of light through non-absorbing disordered dielectrics is usually described by single scattering and multiple scattering, which turns into photon diffusion for many scattering events.  For a wide slab, thickness $L$, the total transmission coefficient $T\left(L\right)$ is given by a ballistic contribution, $T_\text{b}\left(L\right)=e^{-L/\ell}$ with a scattering mean free path $\ell$, a diffusive part $T_\text{d}\left(L\right)$ set by the transport mean free path  $\ell^\ast$ and a crossover term considering the conversion of incident photons to diffusive photons via multiple scattering~\cite{durian1996two,lemieux1998diffusing}. 
 For optically dense samples ($L\gg\ell^\ast$), neglecting surface reflectivity,  the diffusive total transmission coefficient is 
\begin{equation}
T_\text{d}\left(L\right)\simeq \frac{1+z_0}{2z_0+L/\ell^\ast}
\overset{L\gg\ell^\ast} \to
\left(1+z_0\right) \frac{\ell^\ast}{L},
\label{diffT} \end{equation} where $z_0$ is the extrapolation length ratio, a constant of order unity~\cite{ishimaru1978wave,kaplan1993geometric,lemieux1998diffusing,durian1996two}. The scattering  length $\ell$ and the transport mean free path $\ell^\ast$ are linked by the scattering anisotropy parameter $g=\left<\cos \Theta \right>_{d\sigma/d\Omega}$ with $\ell^\ast/\ell=1/\left(1-g\right)$. For very small scatterers, or for the case of stealthy hyperuniformity, the differential scattering cross section becomes zero, $d\sigma/d\Omega \propto \left(\ell\right)^{-1}\equiv 0$, resulting in transparency with $T=1$ independent of $L$~\cite{Leseur2016,graves2008transmission,aubry2020experimental}. 
\newline {\it{Anderson localization of light.--}} Strong Anderson localization (SAL) is an interference effect in multiple scattering of waves leading to exponentially attenuated diffuse transmission through a slab. SC-theory describes SAL by introducing a position-dependent light diffusion coefficient $D\left(z\right)$, where $z$ denotes the distance from the surface of a wide slab~\cite{wolfle2010self,cherroret2008microscopic,cherroret2010transverse}. One needs to solve an implicit equation that contains the average “return probability” which is increased in the presence of SAL and in turn reduces $D\left(z\right)$ from $D\left(0\right)$ at the interface to zero deep inside the medium~\cite{van2000reflection,cherroret2010transverse}. To this end we replace $L/\ell^\ast$ in Eq.~\eqref{diffT} and write: 
\begin{equation}
\frac{L}{\ell^\ast}\to \frac{\tilde L}{\ell^\ast}=\frac{1}{\ell^\ast}\int_{0}^{L}\frac{D_\text{B}}{D\left(z\right)} dz \label{diffLoc} 
\end{equation}
where $D_\text{B}=v_E \ell^\ast/3$ denotes the (Boltzmann) light diffusion coefficient for a speed of light $v_E$. Far from the localization regime $D\left(z\right)\equiv D_\text{B}$ and $\tilde L\equiv L$~\cite{durian1996two,lemieux1998diffusing,carminati2021principles}. For localization in a semi-infinite medium the SC-theory solution is $D_\infty\left(z\right)\simeq D\left(0\right)e^{-2z/\xi}$, where $\xi$ denotes the localization length. By interpolation, for a slab of finite thickness  
Van Tiggelen et al. proposed $D\left(z\right)\simeq D_\infty\left(L/2-\left  |L/2-z\right|\right)$~\cite{van2000reflection,cherroret2010transverse}. 
Due to the mirror symmetry relative to the center of the slab at $z=L/2$ we can take the integral in Eq.~\eqref{diffLoc} from $z\in{\left[0,L/2\right]}$, $\tilde L=2 \int_{0}^{L/2}\frac{D_\text{B}}{D\left(0\right)}e^{2 z/\xi}dz$ and find

\begin{figure}
\includegraphics[width=.9\columnwidth]{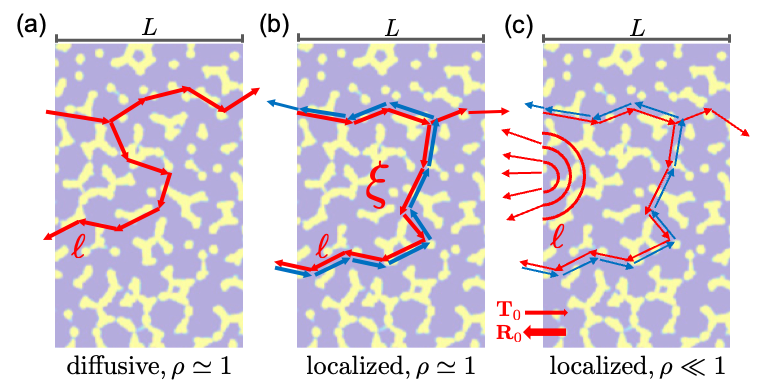}
\caption{\label{fig:1} Transport of light through amorphous hyperuniform silicon networks with localization and bandgap-regimes. Lines indicate scattering paths. $\ell$ denotes the scattering mean free path. (a) random multiple scattering and photon diffusion. (b) Loops of counter propagating paths signaling SAL with a localization length $\xi$ are shown in red and blue color.  (c) PBG-regime with a small but finite DOS.  
The diffusive/SAL part $T_\text{d}\left(L\right)$ is attenuated by a factor $T_\text{0}=\left(1-R_\text{0}\right)$ proportional to the 
normalized photonic density of states (DOS) in the bulk $\rho\left(\nu^\prime\right)$. $R_\text{0}$ denotes the direct reflection factor. The background shows a cross-section of the hyperuniform silicon network displayed in Fig.~\ref{fig:PD}~(a) with a slab thickness $L$.   }\label{fig:Sketch1}
\end{figure}

\begin{equation}
\frac{T_\text{d}\left(L\right)}{\left(1+z_0\right)}=
\frac{1}{2z_0+ \frac{\xi \kern 1pt D_\text{B}}{\ell^\ast D\left(0\right)}\left(e^{L/\xi}-1\right)}\overset{L\gg\xi} \to\frac{\ell^\ast D\left(0\right)
}{\xi \kern 1pt D_\text{B}}e^{-L/\xi}
\label{fulldiffuseT}
\end{equation} 
At the localization transition ('mobility edge'), the full SC-theory, for a finite thickness $L$, predicts a critical power-law decay $T_\text{d}\sim 1/L^2$, instead of an exponential~\cite{van2000reflection}. The onset of this power-law is captured by Eq.~\eqref{fulldiffuseT}, which can be seen by expanding $e^{L/\xi}-1=L/\xi+(L/\xi)^2/2+(L/\xi)^3/6...$, but for thick slabs, $L/\xi\gtrsim 3$, it deviates. In realistic numerical simulations, the systems sizes are limited, and a study of the critical regime is beyond the scope of the present work.
\newline {\it{Direct coherent reflection.--}} Previous studies argued that the transmission probability $T$ ($L\gg\ell^\ast$) is independent of the single scattering angular distribution and only depends on the transport mean free path $\ell^\ast$, as expressed by Eq.~\eqref{diffT}~\cite{ishimaru1978wave,lemieux1998diffusing,durian1996two}. More recent work, driven mainly by the renewed interest in structural coloration, showed the importance of explicitly adding a single scattering reflection term in the presence of correlated disorder, for example, for photonic glasses~\cite{magkiriadou2014absence,hwang2020effects,wang2021recent}. 
Short range order in photonic glasses leads to coherent collective scattering that, for a matching wavelength, results in enhanced single scattering reflection. Compared to structural coloration, the opening of a gap in amorphous photonic materials provides an even stronger, coherent mechanism for direct reflection, fundamentally altering the way scattered photons convert into diffuse photons, Figure~\ref{fig:1}. 
We follow the reasoning of Magkiriadou et al. that the intensity of light directly reflected scales with $\frac{\sigma_\text{dr}}{\sigma}e^{- z/\ell}$, where $\sigma_\text{dr}$ denotes the direct reflection cross section and $\sigma$ the total scattering cross section. The direct reflection from layers close to the surface is higher and the reflected intensity from inside the sample decreases exponentially as the coherent beam attenuates~\cite{magkiriadou2014absence}. Simultaneously, direct reflection reduces the probability density for the creation of diffuse photons within a distance $z$ into the slab which now scales as $\left(1-\frac{\sigma_\text{dr}}{\sigma}\right)e^{- z/\ell}$~\cite{durian1996two,haberko2020transition}. For $\sigma_\text{dr}=\sigma$,  the reflection coefficient $R_\text{0}=\frac{\sigma_\text{dr}}{\sigma} =1$ and all light is coherently reflected in the limit $L\gg\ell$. For $\sigma_\text{dr}<\sigma$ a proportional amount $T_\text{0}=1-R_\text{0}=1-\frac{\sigma_\text{dr}}{\sigma}<1$ can couple to the diffuse up- and downstream of photons. 
Consequently, the total transmission coefficient is lowered to
\begin{equation}T\left(L\gg\ell^\ast\right)=T_\text{0}\times T_\text{d}\left(L\right)\label{T0diff}\end{equation}
We include an approximate expression for $T\left(L\right)$ 
covering the entire range of $L$  in the Supplemental Material, Eq.~(S1).
 \begin{figure}
\includegraphics[width=1\columnwidth]{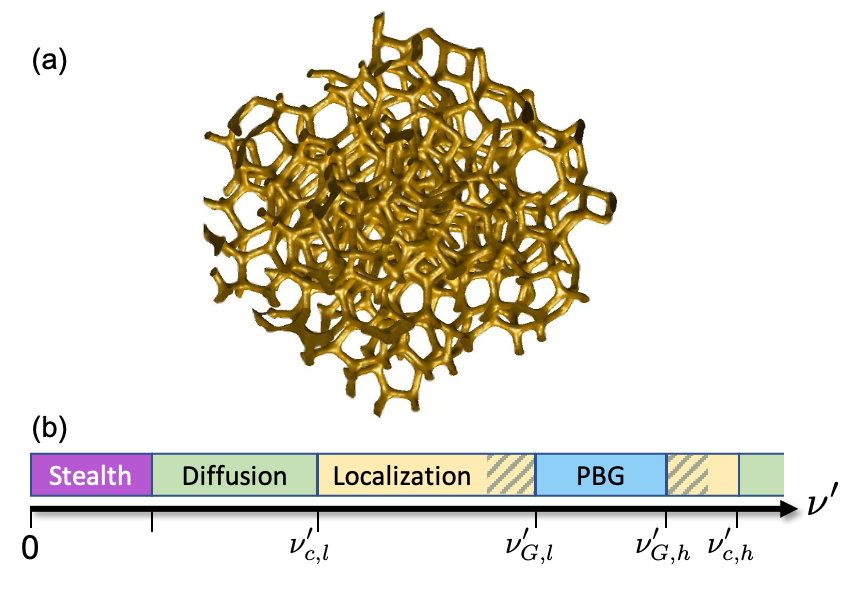}
\caption{\label{fig:PD} (a) Rendering of a three-dimensional hyperuniform silicon network structure in air with a volume filling fraction of $\phi=0.28$ derived from the center positions of randomly close-packed spheres, diameter $a$~\cite{haberko2020transition,Haberko_OPEX_2013},  which also sets the characteristic length scale of short-range spatial correlations~\cite{Cao_PRA_2011}. The refractive index is $n = 3.6$ and the host medium is air, $n_\text{air} = 1$. (b) Transport phase diagram for electromagnetic waves in disordered photonic materials as a function of the dimensionless frequency $\nu^\prime=a/\lambda=\nu(a/v_E)$, adapted from \cite{froufe2017transport}. $\lambda$ denotes the vacuum wavelength and $v_E$ the speed of light. For low frequencies, stealthy hyperuniform materials show transparency. For weak or moderate scattering, light transport is 'diffusive' followed by strong Anderson 'localisation' (SAL) with transitions at $\nu_{c}$ and a band-gap regime ('PBG') around $\nu_{G}$. Closer to the gap the reduced density of states influences localization, shaded areas. The mid-gap frequency is $\nu_{G}\sim 0.50$, in agreement with the Bragg condition in a  corresponding crystal~$\lambda = a/\nu^\prime \sim 2 a$~\cite{haberko2020transition}.}
\end{figure}
 \begin{figure*}
\includegraphics[width=2\columnwidth]{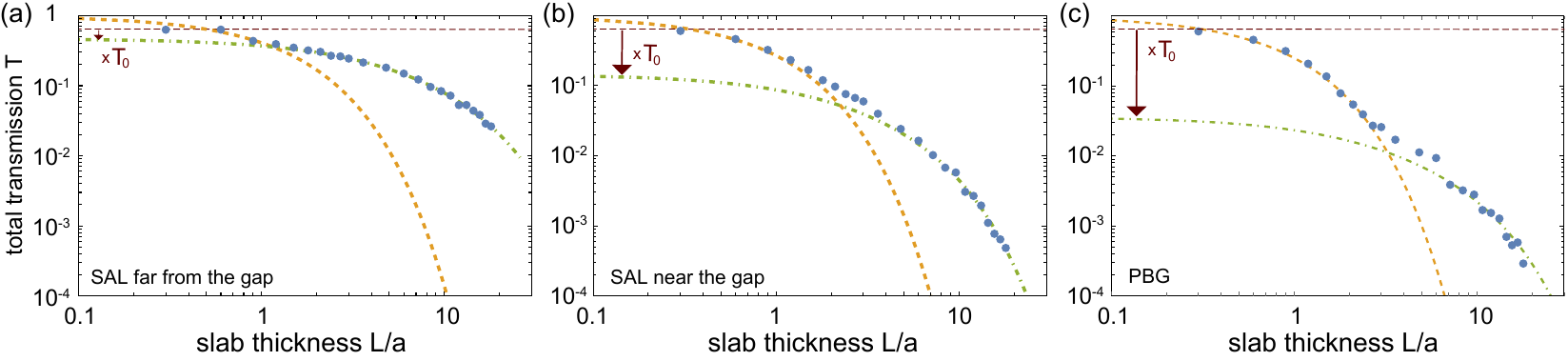}
\caption{\label{fig:TLPlots} Total transmission $T\left(L\right)$ as a function of the reduced slab thickness $L/a$ in log-log representation for three different frequencies $\nu^\prime$ in the localized and bandgap regime. Filled blue circles show the results from FDTD simulations averaged over 6 (thick slabs) to 15 (thin slabs) samples. The dashed orange line shows the fit with $T_\text{b}\left(L\right)=e^{-L/\ell}$ over $L\lesssim a$ which yields $\ell\left(\nu^\prime\right)$. The  dash-dotted green line shows the fit with $T_\text{0}\times T_\text{d}\left(L\right)$ over $7a<L<18a$ which yields $\xi\left(\nu^\prime\right)$ and $T_\text{0}\left(\nu^\prime\right)$. (a) Localization: $\nu^\prime=0.418$, $\ell/a=1.12, \xi/a$= 8 and $ T_\text{0}=0.72\sim 1$. (b) Localization near the PBG: $\nu^\prime=0.462$, $\ell/a=0.76, \xi/a$= 3.6 and $ T_\text{0}=0.22$, (c) Nearly complete PBG: $\nu^\prime=0.471$, $\ell/a=0.72, \xi/a$= 5.1 and $ T_\text{0}=0.055$. The extrapolation length ratio is $z_0=3.25$, taken from~\cite{haberko2020transition}. Horizontal dashed line: $T_\text{d}\left(0\right)=\left(1+z_0\right)/2z_0=0.65$, Eq.\eqref{diffT}.}
\end{figure*}
\newline {\it{Numerical transport simulations.--}} To check the model predictions, Eq.~\eqref{T0diff}, we performed FDTD simulations using the open source MIT Electromagnetic Equation Propagation (MEEP) package on a computer cluster~\cite{oskooi2010meep,haberko2020transition}.  We generate hyperuniform network structures, Fig.~\ref{fig:PD}~(a), using a custom-made code based on a 10,000-particle jammed seed pattern taken from ref.~\cite{song2008phase}, volume filling fraction $\sim 0.64$. Jammed, random close sphere packings display nearly hyperuniform behaviour at large length scales~\cite{Berthier_arXiv_Apr_2015,Froufe_PRL_2016}. All units are given relative to the diameter $a$ of the spheres of the seed pattern. Next, we perform a Delaunay tessellation of the seed pattern. The tessellation divides the pattern into tetrahedra. We connect the centres of mass of the tetrahedra with dielectric rods, creating the desired tetravalent network structure~\cite{Cao_PRA_2011,Muller_Scheffold_Adv_Opt_Mat_2014}. We apply a silicon refractive index $n = 3.6$ and a volume filling fraction of $\phi=0.28$. We cut the digital box into slices to obtain slabs of different thicknesses $L\le18a$, footprint $18a \times 18a$.  In the MEEP simulation, we apply periodic boundary conditions perpendicular to the propagation axis and we add perfectly matched layers (PML) at both ends of the simulation box acting as absorbers. We send a pulse of linearly polarized light and record the Poynting vector on a monitor located behind the structure. The transmission coefficient $T\left(L,\nu^\prime\right)$ is defined as the ratio of the transmitted power to the incoming power.
In total we study twenty three sample thicknesses ranging from $L=0.3 - 18a$, averaged over 6 (thick slabs) to 15 (thin slabs) samples~\cite{haberko2020transition,haberko_jakub_2020_3968424}. 
\newline \indent In Figure~\ref{fig:TLPlots}, we show the FDTD data for the total transmission $T\left(L\right)$ at different frequencies in the localization and bandgap regime. We fit the initial decay for $L\lesssim a$ with $T_\text{b}\left(L\right)=e^{-L/\ell}$ and extract the mean free path $\ell \left(\nu^\prime\right)$. Using this ballistic $\ell \left(\nu^\prime\right)$ we fit the data for thick slabs, $7a<L<18a$ ($L \gg \ell$) using equation Eq.~\eqref{fulldiffuseT} with $z_0=3.25$ from ref.~\cite{haberko2020transition}. For simplicity we assume $\ell^\ast\sim \ell$ and $D_\text{B}/D\left(0\right)\simeq \left (1+2 z_0 /\xi\right)$ in the localized regime~\cite{van2000reflection}. 
 \begin{figure}
\includegraphics[width=.9\columnwidth]{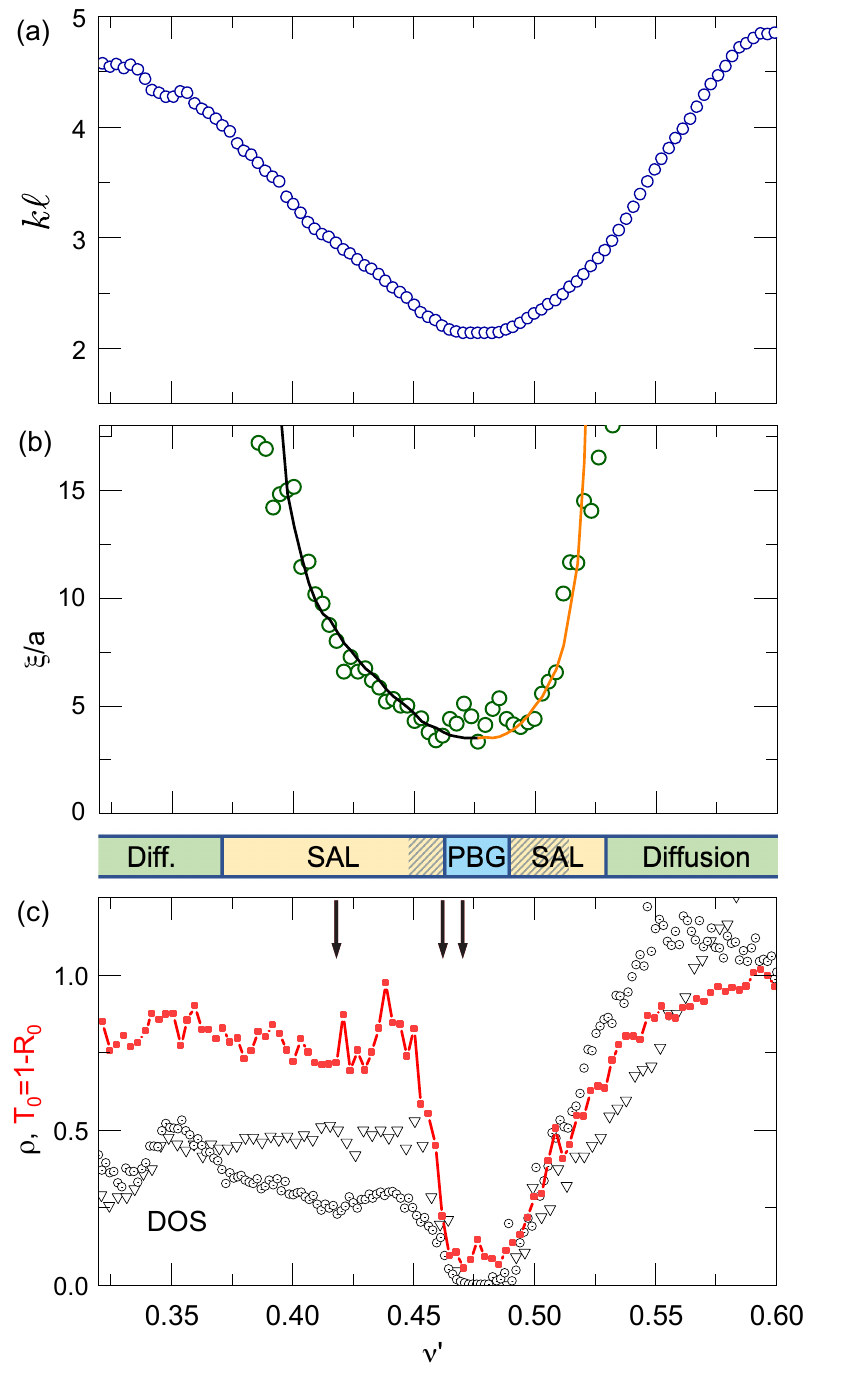}
\caption{\label{fig:FitParamResults} Frequency dependence of the transport parameters determined from the fit to the FDTD data $T\left(L\right)$, as described in Fig.~\ref{fig:TLPlots}.
(a) Reduced mean free path $k\ell=2\pi\ell/\lambda=2\pi\nu^\prime \ell/a$. (b) Localization length $\xi/a$. Lines show the scaling prediction from SC-Theory, $\xi/a\propto\left( k\ell\right)^2/ (1-\left[k\ell/\left(k\ell\right)_c\right]^4)$, with $(k\ell)_{c,l}\simeq 4.1$ (black line) and $(k\ell)_{c,h}\simeq2.85$ (orange line) and a prefactor of order one. The different transport regimes are indicated at the bottom. (c) Red solid squares: $T_\text{0}=1-R_\text{0}$, compared to numerical calculations of the DOS $\rho$ (open symbols). The DOS-data has been reproduced from ref.~\cite{Cao_PRA_2011} (circles) and ref.~\cite{haberko2020transition} (open triangles). The arrows indicate the frequencies for the data shown in Figure \ref{fig:TLPlots}.}
\end{figure}
As shown in Figure~\ref{fig:TLPlots} we find excellent agreement between the fit and FDTD-data.  The attenuation of the incident beam by coherent reflection can become very strong, with $T_\text{0}\to0$, close to and in the PBG. From the fit to $T_\text{d}\left(L\right)$, we extract $\xi\left(\nu^\prime\right)$ and $T_\text{0}\left(\nu^\prime\right)$. 
\newline \indent In Figure~\ref{fig:FitParamResults} we plot the frequency dependence of all parameters obtained by the model fit. The mean free path $\ell$ and thus $k\ell$ decay towards the bandgap and reach a minimal value  $k\ell=\frac{2\pi\ell}{\lambda}\sim 2$ ($\ell/a\sim 0.65$) in the band gap center \cite{haberko2020transition}, Figure~\ref{fig:FitParamResults}~(a). Approaching the gap $\xi/a$ drops and the smallest value we observe is $\xi/a\gtrsim 3$; see also Figure~\ref{fig:TLPlots}(c). van Tiggelen et al. proposed for the $k\ell$ dependence of the localization length  $\xi\propto\left(k\ell\right)^2/\left(1-\left[k\ell/\left(k\ell\right)_c\right]^4\right)$ ~\cite{van2000reflection,cobus2018transverse,skipetrov2018ioffe}. 
    Adjusting $(k\ell)_c$-values we find a lower and upper mobility edge $\nu^\prime\simeq0.37$ at  $k\ell=(k\ell)_c=4.1$ and at $\nu^\prime\simeq0.53$ for $k\ell=(k\ell)_c=2.85$ , Figure~\ref{fig:FitParamResults}~(b). We only consider values $\xi/a<18$ (smaller than the system size). The agreement between theory and data is remarkable. However, we find better agreement with the dimensionless $\xi/a$ compared to the originally suggested $\xi/\ell$. Moreover the expression for $\xi$ does not consider the DOS. In a recent dissertation ~\cite{Monsarrat2022}, Monsarrat derives a slightly different formula for the localisation length from SC-theory that explicitly accounts for the DOS and does not scale with $k\ell^\ast$ in the nominator: $\xi/\ell^\ast\propto \frac{1}{3/\pi-\rho \left(k\ell^\ast\right)^2}$. This expression agrees well with our data for $\xi/\ell$ if we again use $\ell^\ast=\ell$ and replace $k\ell$ by $k \ell/\left(k\ell\right)_c$~\cite{skipetrov2018ioffe}, as shown in the Supplemental Material, Fig.~S2. Note that the influence of the density of states on $\xi$ is small since $\rho<1$ only in a regime where $k \ell/\left(k\ell\right)_c\ll1$.
  \newline {\it{Density of states and coherent reflection.--}}  
It seems plausible that the normalized density of states (DOS)  is responsible for the observed direct reflection since, for a full bandgap, we know that the DOS is zero and $R_\text{0}\equiv 1$. Here we consider the DOS normalized by the density of states of the 'homogeneous' medium~\cite{Cao_PRA_2011,haberko2020transition}. In and near the bandgap the coherent beam's intensity and the $z$-dependent local density of states (LDOS) decay exponentially to their bulk values over a distance of a mean free path $\ell \ll \xi$. In the bulk the mean LDOS is equal to bulk DOS $\rho\left(\nu^\prime\right)$ and thus for $L\gg\ell$ we expect $T_\text{0}\left(\nu^\prime\right)\simeq \rho\left(\nu^\prime\right)$. 
Hasan et al., as well as Skipetrov, argued similarly when discussing finite-size effects in photonic crystals where the incident beam's coherent intensity $T\left( z \right)=1-R\left( z \right)$ and the LDOS decay exponentially. For a crystal the decay length is the Bragg langth $L_\text{B}$~\cite{hasan2018finite,skipetrov2020finite,garcia2011photonic}. Koenderink et al. discussed the attenuation of the coherent beam for the case of disorder in photonic crystals~\cite{koenderink2000enhanced}.
  \newline In Figure~\ref{fig:FitParamResults}~(c) we compare the results for $T_\text{0}\left(\nu^\prime\right)$ from the fit to numerical simulations of the normalized optical density of states (DOS) $\rho\left(\nu^\prime\right)$, published earlier by Hui Cao and co-workers~\cite{Cao_PRA_2011} and by us~\cite{haberko2020transition,haberko_jakub_2020_3968424} using identical design parameters for the hyperuniform dielectric networks shown in Figure~\ref{fig:PD}~(a). We find indeed $T_\text{0}\left(\nu^\prime\right)\sim \rho\left(\nu^\prime\right)$. 
\newline {\it{Origins of the bandgap.--}} Some remarks concerning the bandgap's origins are in order. This work shows that resonant scattering in the presence of correlated disorder leads to destructive interference in the forward direction and enhanced single scattering reflection. Backscattering can also be induced by single scattering resonances~\cite{gomez2012negative}. 
This microscopic picture of preferential backscattering is supported, at least up to the lowest order, by recent diagrammatic calculations~\cite{monsarrat2021pseudogap,saleh2007fundamentals}. 
Zhang-Stillinger and Torquato argued that another structural property is essential for forming a band gap. Uniformity and resulting bounded hole sizes (empty regions) prevent deep penetration of unscattered photons into the bulk of the material~\cite{C7SM01028A,ghosh2018generalized}.  
They say that such structural "rigidity" confers novel physical properties to disordered systems, including the desired band gap.  Qualitatively, stealthy hyperuniformity can provide both mechanisms. Stealthiness implies bounded hole sizes~\cite{C7SM01028A} and suppresses scattering at scattering angles $\vartheta<\vartheta_c$ ($q_\vartheta=2k\sin \left(\vartheta/2\right)<q_c$) through collective scattering and destructive interference~\cite{Leseur2016,torquato2015ensemble}.
For a quantitative microscopic assessment, however, higher-order scattering loops, beyond the collective scattering approximation, must be taken into account~\cite{monsarrat2021pseudogap}.
\newline {\it{Summary and Conclusion--}} In this work, we study the transmission of light through hyperuniform, high refractive index networks using numerical simulations and theory. We propose adding a direct reflection term to the theoretical model describing light transmission through optically dense amorphous photonic materials. Combined with the results of the self-consistent theory of localization (SC-theory) for a semi-infinite medium, we derive a simple, closed-form analytical expression for the total transmission coefficient of optically dense slabs $T_\text{d}\left(L\right)$ ~\cite{wolfle2010self,cherroret2010transverse,van2000reflection}. Our model captures the optical transmission behavior between localization and the bandgap.  Moreover, we rationalize that near and inside the gap, the reduced density of states is responsible for the coherent reflection and the attenuation of the coupling of the incident light beam to diffuse and localized transport. The quantitative agreement of our theory with numerical simulations suggests that it could be of considerable value for experimental studies.
Moreover, this study shows that for an amorphous PBG material in the gap, the scattering mean free path $\ell$ is equivalent to the Bragg length in a photonic crystal~\cite{Joannopoulos_book,garcia2009strong}. This observation is essential, and the behavior is different from disordered photonic crystals, where the Bragg length is given by the periodically repeating environment and the scattering length by the degree of disorder~\cite{koenderink2000enhanced,conti2008dynamic,garcia2011photonic,aeby2021scattering}.  

\acknowledgments We thank Sergey Skipetrov and Arthur Goetschy for insightful discussions. The Swiss National Science Foundation supported this work through the National Centre of Competence in Research 'Bio-Inspired Materials', \# 182881 (FS and LFP), projects number \# 149867 (FS) and \# 97146 (LFP).

\clearpage
\section*{Supplemental Material}
\section*{Frank Scheffold et al.} 
\vspace{1mm}
\setcounter{equation}{0}
\setcounter{figure}{0}
\setcounter{table}{0}
\setcounter{page}{1}
\makeatletter
\renewcommand{\theequation}{S\arabic{equation}}
\renewcommand{\thefigure}{S\arabic{figure}}

\title{Light transport through amorphous photonic materials with localization and bandgap regimes \\ -------- Supplemental Material -------- }

{\bf{Total transmission over the entire range of slab thicknesses.}} Durian's two-stream theory for the propagation of light through a randomly scattering slab provides a simple expression for the total transmission coefficient covering the entire range of slab thicknesses from $L=0$ to thick slabs~\cite{durian1996two,lemieux1998diffusing}. The theory was later reformulated and generalized in terms of a telegrapher equation ~\cite{lemieux1998diffusing} but the expression we use here is the same in both works. In his two-stream theory, Durian describes scattering photon transport by two concentration currents, an up- (forward) and down- (backward) stream. The theory takes into account ballistic, diffusive scattering and the cross-over regime where the incident photons are converted into diffuse photons. For classical scattering, e.g. in the absence of PBG and SAL, the theory describes transport in three dimensions more accurately than diffusion theory. 
Notably, it covers the cross-over regime to thin slabs and reproduces the expected $T\left(L=0\right)=1$, neglecting (Fresnel) surface reflectivity at normal incidence (typically of a few percent). We can generalize the results by Durian ad-hoc by multiplying the multiple scattering term, Eq.~$\left(14\right)$ in ref.~\cite{durian1996two}, by $T_0$. Using Eq.~\eqref{DurianT}, again with $\ell^\ast \simeq \ell$, we can describe the FDTD transport data for $T\left(L/a, \nu^\prime\right)$ over the entire range of $L/a$ as shown in Figure~\ref{fig:FitTall}. The ad-hoc generalization appears to work well, but we note that to consistently merge SC theory with the theory by Durian, additional corrections appear as discussed in \cite{haberko2020transition}.
A full theoretical treatment is beyond the scope of the present work and shall be addressed in the future.
\begin{widetext}
\begin{equation}
T\left(L\right) =  T_\text{b}+T_0\times T_\text{d}\simeq e^{-L/\ell}\\+T_0\left[\frac{\left(1+z_0\right)}{2 z_0 + \tilde L/\ell^\ast}\left(1-e^{-L/\ell}\right)
-\frac{\tilde L/\ell^\ast}{2 z_0 + \tilde L/\ell^\ast}e^{-L/\ell}\right]\label{DurianT}
\end{equation}
\begin{figure*}[h]
\includegraphics[width=.95\columnwidth]{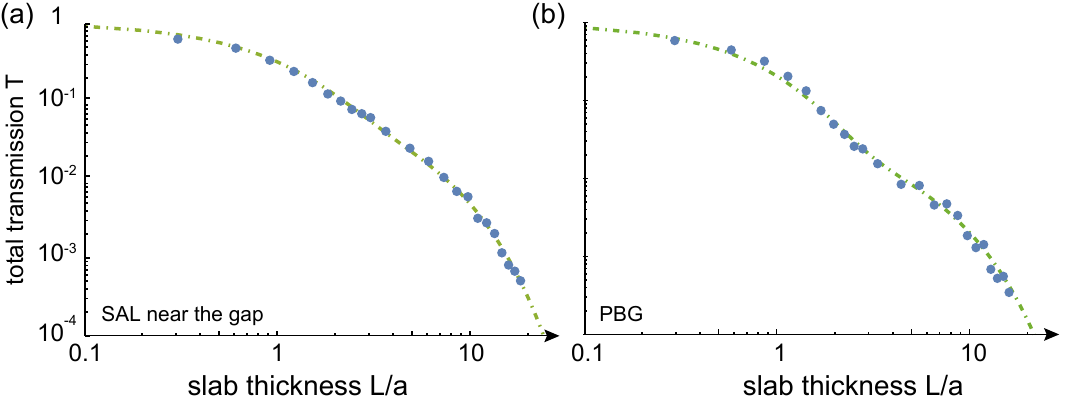}
\caption{\label{fig:FitTall} Total transmission $T\left(L,\nu^\prime\right)$ as a function of the reduced slab thickness $L/a$ in log-log representation for two different frequencies $\nu^\prime$ in the localized and bandgap regime. Symbols denote the results from FDTD simulations averaged over 6 (thick slabs) to 15 (thin slabs) samples. The dash-dotted green line shows the curve fitted with Eq.~\eqref{DurianT} over $7a<L<18a$. a) $\nu^\prime=0.462$, $\ell/a=0.75$, $\xi/a$= 3.6 and $1-R_0=0.22$, b) $\nu^\prime=0.474$, $\ell/a=0.62$, $\xi/a$= 4.5 and $ 1-R_0=0.09$. The extrapolation length ratio is $z_0=3.25$, taken from ref.~\cite{haberko2020transition}.} 
\end{figure*}
\end{widetext}

{\bf{Localisation length for the 3D case.}}
In a recent dissertation, Monsarrat proposes an expression for the localization length derived from SC-theory in 3D ~\cite{Monsarrat2022}: 
\begin{equation}
    \xi/\ell^\ast = \frac{3}{2}\frac{1}{3/\pi-\rho \left(k\ell^\ast\right)^2} \label{MonsEq}\end{equation} 
Eq.~\eqref{MonsEq} explicitly considers the normalized DOS ($\rho$) and the localization length is expressed in units of $\ell^\ast$. 
For a comparison to the values of $\xi/a$ shown in Figure~4~(b), we again use $\ell^\ast=\ell$ and replace $k\ell$ by $k \ell/\left(k\ell\right)_c$. Moreover, we assume $\rho\equiv T_0$ for the density of states, and thus all parameters are defined from the fit of Eq.~(3 ) to the FDTD data. 
\begin{equation}
    \xi/\ell  \propto \frac{1}{3/\pi-T_0 \left(k\ell\right)^2} \label{MonsEq2}\end{equation} 
We note that the influence of the density of states on $\xi$ is small since $\rho<1$ only in a regime where $k \ell/\left(k\ell\right)_c\ll1$. 
\begin{figure}[h]
\includegraphics[width=1\columnwidth]{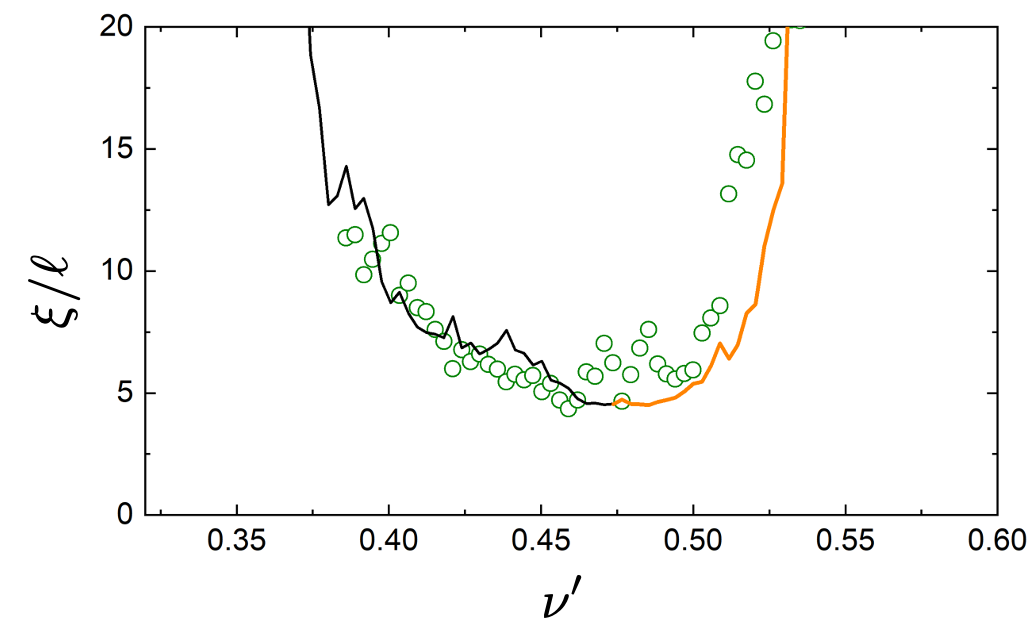}
\caption{\label{fig:xifit22B} Localization length in units of the mean free path $\xi/\ell$ as a function of the dimensionless frequency $\nu^\prime=a/\lambda$ (only data with $\xi/a<18$ shown). Lines show the prediction from SC-Theory in three dimensions according to Monssarat~\cite{Monsarrat2022}, Eq.\eqref{MonsEq2} with $(k\ell)_{c,l}\simeq 4.1$ (black line) and $(k\ell)_{c,h}\simeq2.85$ (orange line), input parameters from the FDTD fit and a prefactor on the order of one. }
\end{figure}
\newline We use the same values of $(k\ell)_{c,l}\simeq 4.1$ (black line) and $(k\ell)_{c,h}\simeq2.85$ as in Figure~4.
In Figure~\ref{fig:xifit22B} we compare the prediction by Eq.~\eqref{MonsEq2} to the data for $\xi/\ell$ and find excellent agreement.
\smallskip

\end{document}